\documentclass[aps,prx,amsmath,amssymb,amsfonts,lengthcheck,twocolumn,superscriptaddress]{revtex4-2}

\usepackage[T1]{fontenc}
\usepackage{graphicx}
\usepackage{subfigure}
\usepackage{amsmath, amsthm, amssymb, amsfonts}
\usepackage{mathtools}
\usepackage{verbatim}
\usepackage{dcolumn}
\usepackage{bm}
\usepackage{epsf}
\usepackage{color}
\usepackage[colorlinks=true,citecolor=blue,linkcolor=blue,urlcolor=blue]{hyperref}
\usepackage{xcolor}
\usepackage{physics}
\usepackage{dsfont}
\usepackage{tikz}
\usepackage{multirow}
\usepackage{cleveref}
\usepackage{siunitx}
\usepackage{xurl}

    \newcommand{\cbqty}[1]{\left\{#1\right\}}

\begin{document}


\title{Thermodynamic significance of QUBO encoding on quantum annealers}

\author{Emery Doucet}
\thanks{These authors contributed equally}
\email{emery.doucet001@umb.edu}
\affiliation{Department of Physics, University of Massachusetts, Boston, Boston, MA 02125, USA}
\affiliation{Department of Physics, University of Maryland, Baltimore County, Baltimore, MD 21250, USA}
\affiliation{Quantum Science Institute, University of Maryland, Baltimore County, Baltimore, MD 21250, USA}

\author{Zakaria Mzaouali}
\thanks{These authors contributed equally}
\email{z.mzaouali@extern.fz-juelich.de}
\affiliation{Institut für Theoretische Physik, Universität Tübingen, Auf der Morgenstelle 14, 72076 Tübingen, Germany}
\affiliation{Jülich Supercomputing Centre, Institute for Advanced Simulation, Forschungszentrum Jülich, Wilhelm-Johnen-Straße, Jülich, 52428, Germany.}

\author{Reece Robertson}
\affiliation{Department of Computer Science and Electrical Engineering, University of Maryland, Baltimore County, Baltimore, MD 21250, USA}
\affiliation{Department of Physics, University of Maryland, Baltimore County, Baltimore, MD 21250, USA}
\affiliation{Quantum Science Institute, University of Maryland, Baltimore County, Baltimore, MD 21250, USA}

\author{Bartłomiej Gardas}
\affiliation{Institute of Theoretical and Applied Informatics, Polish Academy of Sciences, Ba{\l}tycka 5, Gliwice, 44-100, Poland}

\author{Sebastian Deffner}
\affiliation{Department of Physics, University of Maryland, Baltimore County, Baltimore, MD 21250, USA}
\affiliation{Quantum Science Institute, University of Maryland, Baltimore County, Baltimore, MD 21250, USA}
\affiliation{National Quantum Laboratory, College Park, MD 20740, USA}

\author{Krzysztof Domino}
\email{kdomino@iitis.pl}
\affiliation{Institute of Theoretical and Applied Informatics, Polish Academy of Sciences, Ba{\l}tycka 5, Gliwice, 44-100, Poland}

\begin{abstract}
Quadratic unconstrained binary optimization (QUBO) is the standard interface to quantum annealers, yet a single constrained task admits many QUBO encodings whose penalty choices reshape the energy landscape experienced by hardware. We study a Job Shop Scheduling instance using a two-parameter family of encodings controlled by penalty weights $p_{\rm sum}$ (one-hot/sum constraints) and $p_{\rm pair}$ (precedence constraints). Sweeping $(p_{\rm sum},p_{\rm pair})$, we observe sharp transitions in feasibility and solver success across classical annealing-inspired heuristics and on a D-Wave Advantage processor. Going beyond solution probability, we treat the annealer as an open thermodynamic system and perform cyclic reverse-annealing experiments initialized from thermal samples, measuring the stochastic processor energy change. From the first two moments of this energy change we infer lower bounds on entropy production, work, and exchanged heat via thermodynamic uncertainty relations, and corroborate the observed trends with adiabatic master equation simulations. We find that the same encoding transitions that govern computational hardness also reorganize dissipation: weak penalties generate low-energy infeasible manifolds, while overly strong penalties suppress the effective problem energy scale and increase irreversibility, reducing the thermodynamic efficiency. Our results establish QUBO penalties as thermodynamic control knobs and motivate thermodynamics-aware encoding strategies for noisy intermediate-scale quantum annealers.
\end{abstract}

\maketitle
\section{Introduction}
Quantum annealing (QA) provides a hardware-native route to solving discrete optimization problems by encoding them into the low energy state of an Ising Hamiltonian and approximately preparing that ground state through a driven open system evolution~\cite{lucas2014ising, glover2019qubo, domino2025baltimore, Domino2023, Robertson2025, hanussek2025solvingquantuminspireddynamicsquantum, ewa2026}.
Over the last decade, programmable superconducting annealers have grown to thousands of coupled qubits with improving connectivity and control, enabling systematic studies of combinatorial workloads under finite temperature, noise, and calibration constraints~\cite{johnson2011qa}.
A broad range of practical tasks--including scheduling, routing, and resource allocation--can be expressed in the standard quadratic unconstrained binary optimization (QUBO) form and mapped to the Ising model implemented on these devices~\cite{stollenwerk2019flight, carugno2022jobshop, Domino2023}.
This practical expressivity has motivated extensive benchmarking efforts and fueled interest in whether QA can yield computational advantage for optimization~\cite{PawlowskiClosingGap,Lidar2025, ChandaranaKipuAdvantage}.

At the same time, it is increasingly clear that runtime-based narratives of ``quantum advantage'' in optimization are difficult to make robust on present-day devices.
Reported speedups can depend sensitively on the performance metric, the choice of classical baseline, and, critically, on how one accounts for end-to-end overhead such as programming, readout, thermalization, compilation/transpilation, and hybrid post-processing.
Recent reassessments emphasize that on NISQ hardware the separation between ``pure compute'' time and overhead is often not experimentally clean, so omitting these contributions can systematically bias runtime comparisons \cite{tuziemski2025recentquantumruntimedisadvantages, mr2n-qqrb}.
This does not diminish the scientific value of QA as a controllable many-body open quantum system, but it motivates broadening the notion of ``performance'' beyond time-to-solution alone~\cite{R_nnow_2014}.

One natural complementary axis is energetic cost: how much work must be invested, how much heat is dissipated, and how irreversible is the physical process that produces candidate solutions~\cite{hauke2020perspectives, campbell2025roadmap, alexia2022, Mzaouali2021, Stevens_2025, Deffner_2021}.
Unlike idealized asymptotic runtime, these thermodynamic quantities are intrinsic to the device-level dynamics and directly reflect the interplay between driving, dissipation, and the encoded energy landscape~\cite{gardas2018quantum}.
Thermodynamic observables can therefore serve as a hardware-grounded diagnostic of why certain problem encodings succeed or fail, and they can inform encoding strategies aimed at reducing dissipation and improving robustness to noise.
In this work we adopt the viewpoint that quantum computers are thermodynamic machines characterized by work, heat, entropy production, and thermodynamic efficiency inferred from experimentally accessible statistics~\cite{Campisi2021, Buffoni2020, smierzchalski2024efficiency}.

A central practical challenge in QA is that many real-world tasks are naturally formulated as constrained optimization problems, whereas QUBO is unconstrained~\cite{glover2018tutorial}.
Although other methods exist (e.g., \cite{SCOOP}), constraints are typically enforced by adding penalty terms with tunable weights.
Even for a fixed problem, there is rarely a unique QUBO: distinct choices of penalty structure and penalty magnitude can encode the same feasible optimum while producing markedly different spectra, degeneracies, and barrier structures~\cite{Karimi_2017}.
Since annealers operate at finite temperature with control noise and limited coefficient ranges, these encoding choices can strongly affect both (i) the probability of recovering feasible near-optimal solutions and (ii) the dissipative cost of the annealing dynamics~\cite{gusmeroli2022expedis, hrga2021madam}.
Thus, QUBO design is not merely a mathematical preprocessing step but a physically consequential co-design problem between algorithmic encoding and hardware dynamics~\cite{apolloni1989quantum, kadowaki1998quantum, farhi2014quantum}.

Here we make this connection concrete using a Job Shop scheduling problem as a representative constrained workload~\cite{pinedo2012scheduling}.
We introduce a two-parameter family of QUBO encodings controlled by penalty weights \(p_{\rm sum}\) (one-hot/sum constraints) and \(p_{\rm pair}\) (precedence constraints), and we map the resulting encoding space to both solver performance and thermodynamic signatures.
On the computational side, we identify sharp regime boundaries separating feasible and infeasible ground states and quantify how these boundaries manifest across classical annealing-inspired solvers and on a D-Wave Advantage processor.

On the thermodynamic side, we extract bounds on entropy production, work, and heat from reverse-annealing experiments using thermodynamic uncertainty relations~\cite{TUR1,TUR2,TUR3}, and we complement these measurements with open-system simulations of the annealing dynamics.
We show that encoding-induced rearrangements of the energy landscape produce distinct thermodynamic regimes, and that the encoding choices associated with ``hard'' optimization behavior also correlate with increased dissipation and reduced thermodynamic efficiency.
This establishes thermodynamics as an operational lens for understanding and improving QUBO encodings on noisy intermediate-scale quantum annealers.

\begin{figure*}
    \centering
    \includegraphics[width=\linewidth]{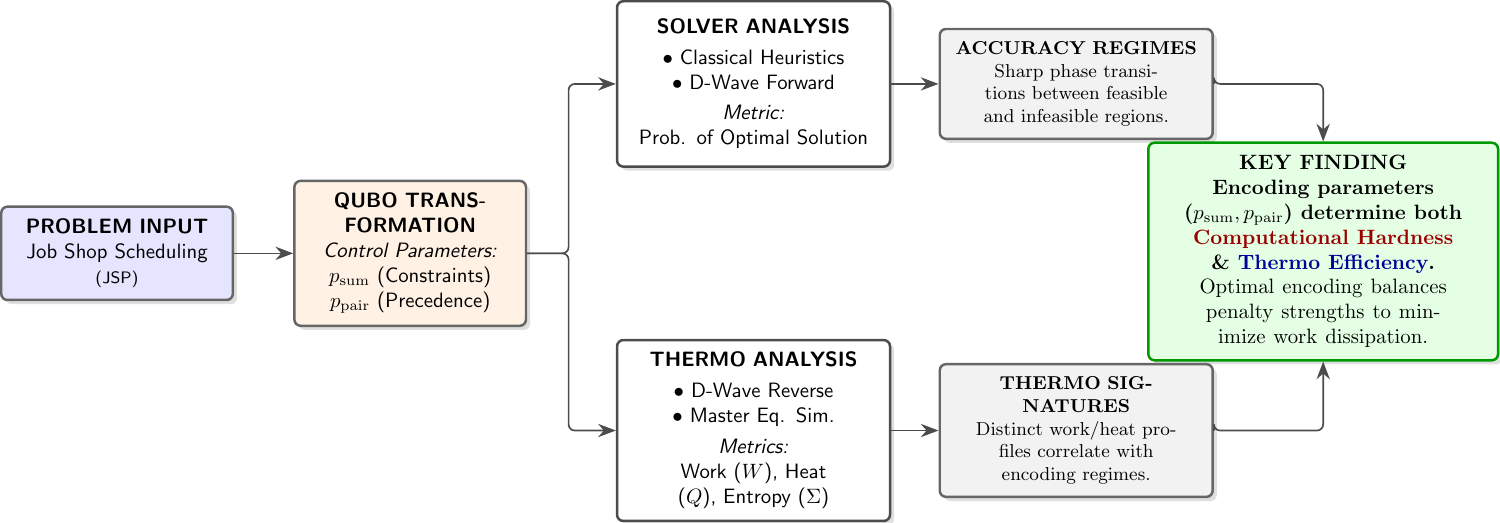}
    \caption{\textbf{Workflow of the paper}. The Job Shop Scheduling Problem is mapped to a QUBO Hamiltonian characterized by tunable penalty parameters $psum$ (for one-hot constraints) and $ppair$ (for precedence constraints). We perform a parallel analysis of this encoding: assessing computational accuracy via solution probability (top branch) and thermodynamic cost via work $\langle W \rangle $, heat $\langle Q \rangle$, and entropy production $\langle \Sigma \rangle$ (bottom branch). The results demonstrate that phase transitions in the problem's computational hardness coincide with distinct thermodynamic signatures, establishing that optimal encoding parameters are those that minimize work dissipation during the annealing schedule.}
    \label{fig0}
\end{figure*}

The outline of this manuscript is as follows.
In Sec.~\ref{sec:JobShopProblem}, we introduce the notion of a Job Shop scheduling problem (JSP), a very important class of scheduling problem which we use as a test case throughout this work. 
In Sec.~\ref{sec:ClassicalSolvers}, we apply several classical simulated-annealing-type solvers to our problem and show how the penalty parameter space is divided into distinct regions. 
Section~\ref{sec:ThermodynamicsOfAnnealing} gives a description of the thermodynamics of quantum annealing, in particular showing how experiments using reverse annealing can extract thermodynamic quantities such as total work and heat. 
Also given is a description of how these quantities can be extracted from open-system simulations of the quantum annealing process.
Our experimental and numerical results are shown in detail in Sec.~\ref{sec:Results}, followed by our conclusion in Sec.~\ref{sec:Conclusion}.

\section{Job Shop problem}
\label{sec:JobShopProblem}

The Job Shop scheduling problem is a fundamental optimization problem in which multiple jobs undergo a sequence of operations on specific machines under strict precedence and resource constraints~\cite{pinedo2012scheduling,Blazewicz2007}. Its practical relevance is demonstrated extensively in manufacturing and production systems, where efficient sequencing of machining and assembly tasks is critical for throughput and cost reduction~\cite{JainMeeran2003}. JSP principles also underpin planning in complex industrial environments, such as semiconductor fabrication, where thousands of operations must be coordinated on limited equipment~\cite{Monch2013}. Similar structured sequencing constraints occur in aircraft maintenance operations, where tasks must be executed on limited facilities and in prescribed orders to ensure operational reliability~\cite{Moser2019}.

Beyond manufacturing, Job Shop-type formulations appear naturally in transportation systems, including railway traffic and rescheduling problems, where trains must traverse shared infrastructure while respecting ordered procedures and resource conflicts~\cite{HansenPachl2014,Cacchiani2014,Domino2023}. 
The problem is even more interesting in real-time railway rescheduling. When disruptions occur, tasks such as reallocating tracks, reordering trains, adjusting meet–pass decisions, or restoring circulation plans require resolving conflicts akin to dynamic Job Shop scheduling under uncertainty. Recovery optimization methods explicitly draw on sequencing and resource-allocation principles from JSP to ensure operational feasibility and minimize delay propagation across the network~\cite{Cacchiani2014}. Here, the quantum devices are good candidates to handle such problems, as they are, by nature, noisy and prone to errors~~\cite{domino2025baltimore, NISQ_Grovers}. Henceforth, assessing the nature of the noise of such (stochastic) solvers is crucial in handling practical scheduling or rescheduling problems under uncertainty.

Let us now introduce the Job Shop problem from the mathematical point of view of scheduling theory~\cite{pinedo2012scheduling}.
We consider a Job Shop problem with release ($r_j$) and deadline ($d_j$) constraints, where the objective is the total weighted tardiness $\sum_j w_j T_j$, namely:
\begin{equation}
    J_m | r_j d_j | \sum_j w_j T_j.
\end{equation}
Each job has its own schedule corresponding to the sequence of machines it must be processed on, and each machine can process one job at a time (i.e. sometimes jobs have to wait in the buffer before the machine is free). 
Release constraints are the earliest possible times jobs may be submitted to the first machine in their schedule, and deadline constraints are the times at which the job must have been processed with all requisite machines. 
Beyond the deadlines, each job should be performed as soon as possible to minimize the weighted tardiness objective.
This type of scheduling problem is
generally NP-hard~\cite{baptiste2001constraint}.

\subsection{QUBO Formulation}

The standard methodology used when solving a Job Shop problem classically involves representing the problem as an integer linear programming (ILP) problem~\cite{pinedo2012scheduling} (For details on this approach, see Appendix.~\ref{app:ILPEncoding}).
For quantum solvers however, the most natural representation used to solve optimization problems is as a quadratic unconstrained binary optimization (QUBO) problem.

Following Ref.~~\cite{venturelli2016job}, we may derive the QUBO directly from the Job Shop problem using a vector $\vec{x}$ of binary decision variables $x_{j,m,t}$ defined to be $1$ when the schedule is such that job $j$ completes on machine $m$ at time $t$ and $0$ otherwise. 
A detailed construction of the QUBO associated with our Job Shop problems is provided in Appendix~\ref{app:QUBOEncoding}.
In summary, the various constraints in the original problem are represented as penalty terms of the form
\begin{align}
    p_{\rm sum}\sum_{m,j}\pqty{\sum_{t,t': t \neq t'} x_{m,j,t}x_{m,j,t'} - \sum_{t} x_{m,j,t}^2}
    ,
    \label{eq::psum_main}
\end{align}
and
\begin{align}
    p_{\rm pair}\sum_{i<i'}\pqty{x_i x_{i'} + x_{i'} x_i}
    ,
    \label{eq::ppair_main}
\end{align}
included in the objective function of the QUBO. 
Combined with the original weighted tardiness cost function of the original problem, the objective function is a purely linear contribution to the QUBO,
\begin{align}
    \text{objective}(\vec{x})
    = 
    \sum_j \sum_{t} w'_j\,t\,x_{j, m_{j, \text{end}}, t} - \text{offset}
    .
    \label{eq::objective_main}
\end{align}
For details on the bounds on these sums and on exactly which terms appear, see Appendix~\ref{app:QUBOEncoding}.
What is important here is the structure of the terms.

\begin{table}[t]
    \centering
    \begin{tabular}{ c c c c c c c}
        \hline\hline
        \#\,qubits & \#\,jobs & \#\,machines & $r_{1,2, \ldots}$ & $d_{1,2, \ldots}$ & $w_{1,2, \ldots}$ & opt. obj.\\ 
        \hline
        $4$ &  $2$ & $1$ & $1,1$ & $3,3$ & $1, 0.5$ & $0.5$\\
        $5$ & $2$ & $1$ & $1,1$ & $3,4$ & $1, 0.5$ & $0.25$\\
         $6$ &  $2$ & $2$ & $1,1$  & $4,3$ & $1, 0.5$& $0.5$ \\
        $8$ &  $2$ & $3$ & $1,1$  & $4,4$ & $1, 0.5$& $0.5$ \\
        $10$ &  $2$ & $2$ & $1,2$  & $7,7$ & $1, 1$ & $0.5$\\
        \hline\hline
    \end{tabular}
    \caption{Job Shop scheduling benchmark instances used in this work.
For each instance we report the number of binary decision variables in the corresponding QUBO formulation (denoted as ``\# qubits''), the number of jobs $J$ and machines $M$, the job-specific release times $r_j$, deadlines $d_j$, and tardiness weights $w_j$ (listed in order of job index $j=1,\ldots,J$). The final column gives the optimal objective value of the original constrained Job Shop scheduling problem, defined as the minimum total weighted tardiness $\sum_j w_j T_j$, and serves as the ground-truth reference for evaluating solver performance.}
    \label{tab:instances}
\end{table}

The Job Shop instances (each job with a predefined sequence of machines) are presented in Tab.~\ref{tab:instances}. Various values of $w_j$ yield various job priorities. 
\subsection{Parameter regimes}

Structurally, the Job Shop problem we have defined in the previous section remains exactly the same as we vary the penalty parameters $p_{\rm sum}$ and $p_{\rm pair}$ used to embed the original constrained optimization problem in QUBO form.
That said, there are distinguished points in the parameter space where important changes occur in the nature of the solution to the QUBO or to the spectrum of the corresponding Ising model.
The difficulty in solving the QUBO may potentially be quite different in the different regions of the parameter space, and hence these transitions could be visible in the performance metrics used to benchmark QUBO solvers.

\begin{figure}[t]
    \centering
    \includegraphics[width=\columnwidth]{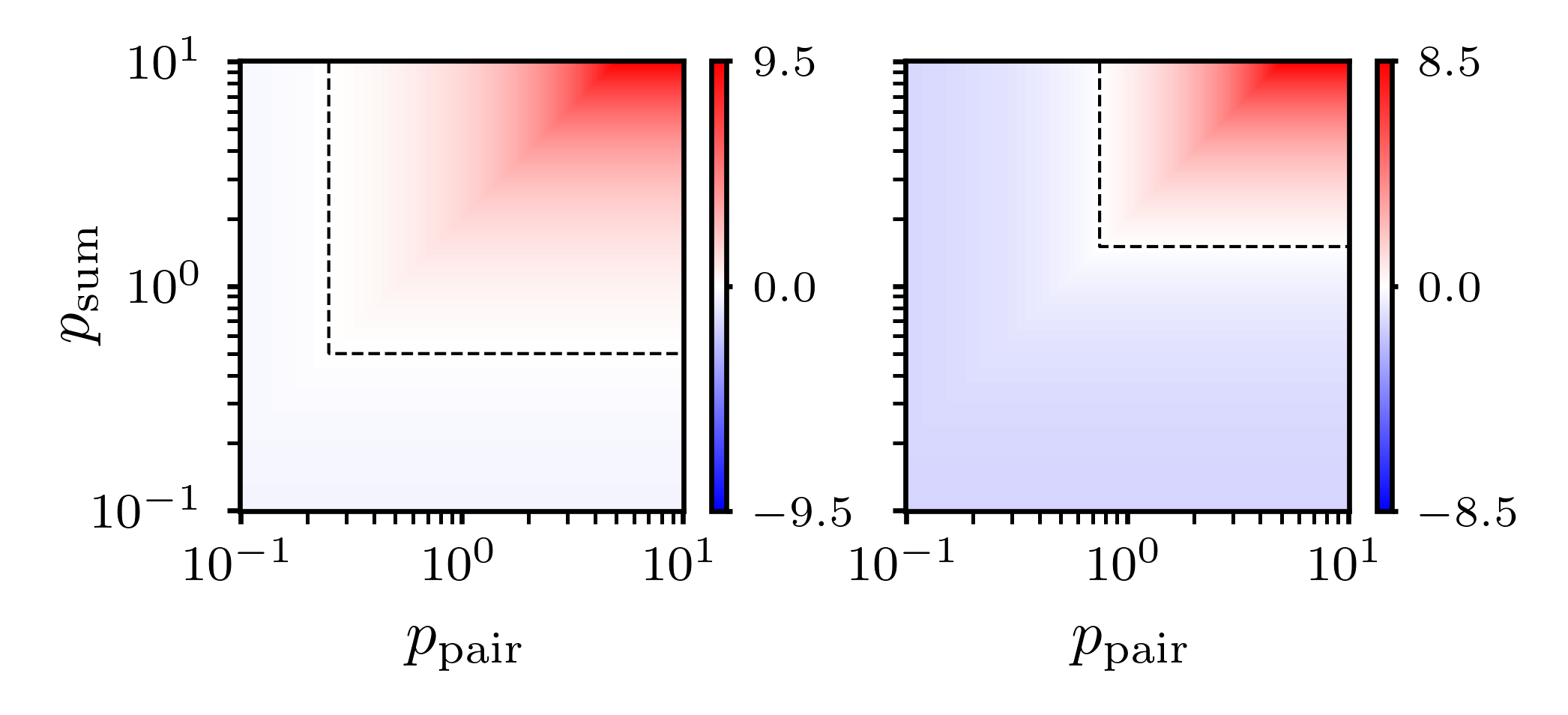}
    \caption{The left plot shows the difference between the best feasible solution and the best infeasible solution. Here the dashed line marks the region where the solution to the QUBO is a valid solution to the original Job Shop problem. The right plot shows the difference between the worst feasible and best infeasible solutions, with the border between the ``split'' and ``unsplit'' regimes marked in black.}
    \label{fig:Splitness6QB}
\end{figure}

As an example, take the problem instance with 8 variables. 
For this case, the solution to the QUBO is not a feasible solution to the original Job Shop problem due to constraint violations unless the penalty values are sufficiently large, $p_{\rm sum} > 0.5$ and $p_{\rm pair} > 0.25$.
This is demonstrated in the left pane of Fig.~\ref{fig:Splitness6QB}, where the difference in objective between the QUBO solution and the best feasible solution is plotted showing the clear separation of the parameter space into two regions. 
Another demarcation not directly related to the optimal solution is plotted in the right pane of Fig.~\ref{fig:Splitness6QB}, showing the difference between the best infeasible solution and the worst feasible solution. 
Again we find the parameter space to be split into two regions, one where all feasible solutions have lower objective values than all infeasible solutions and one where they are mixed. 
Intuitively this might be expected to change the difficulty in achieving a high-quality solution for annealing-based QUBO solvers.
Segregating infeasible solutions at higher energies may reduce the probability they are explored in the annealing process, or the increased separation may reduce the likelihood that high-lying states decay effectively.

One interesting detail of these plots is that they are not symmetric between $p_{\rm sum}$ and $p_{\rm pair}$. 
This is a consequence of fundamental difference between how the two penalty terms interact with the objective function, and we expect that this asymmetry will manifest in the difficulty in solving the problem and hence in the efficacy and thermodynamic efficiency of solvers.

To explain, observe that only negative terms arise from the diagonal part of Eq.~\eqref{eq::psum_main}.
These may be overruled by terms coming from the objective function in Eq.~\eqref{eq::objective_main} if the $p_{\rm sum}$ parameter is too small. 
If there are no negative terms at all in the general QUBO formulation, then the solution is trivial to find~\cite{kochenberger2014unconstrained} and hence should require little effort from the solver. 
Conversely, the suggestion is that more and larger negative terms should make the problem more difficult to solve.
Therefore, we expect that $p_{\rm sum}$ should have a larger impact on solver performance (and other proxies for difficulty of the QUBO) than the other penalty terms proportional to $p_{\rm pair}$. 

Physically, annealer devices work with the Ising problem associated to a QUBO problem.
The binary decision variables $\vec{x}$ are replaced with spins $\vec{s} = 2\vec{x}-1$ taking values in $\cbqty{-1,1}$, producing an energy function to be minimized,
\begin{equation}
E_{\text{Ising}}(\vec{s})
    = \sum_i h_i s_i
    + \frac{1}{2}\sum_{i, i': i\neq i'} J_{ii'} s_i s_{i'}
    \label{eq::Ising_form}
\end{equation}
where the local fields $h_{i}$ are built from the linear and quadratic part of the binary optimization problem (by linear transformation), and the couplings $J_{ii'}$ from the quadratic part~\cite{lucas2014ising}.
Zero values of local fields yield degeneracy in the ground state. Henceforth, we expect some linear pattern for small $p_{\text{sum}}$ and $p_{\text{pair}}$ on space diagrams. Such patterns are expected to pinpoint some positive $p_{\text{sum}}$ at zero $p_{\text{pair}}$ due to the presence of the positive objective diagonal terms in Eq.~\eqref{eq::objective_main}, i.e., in the original QUBO form.

\section{Behavior of Classical Solvers}
\label{sec:ClassicalSolvers}

\begin{figure}[t]
    \centering
    \includegraphics[width=\columnwidth]{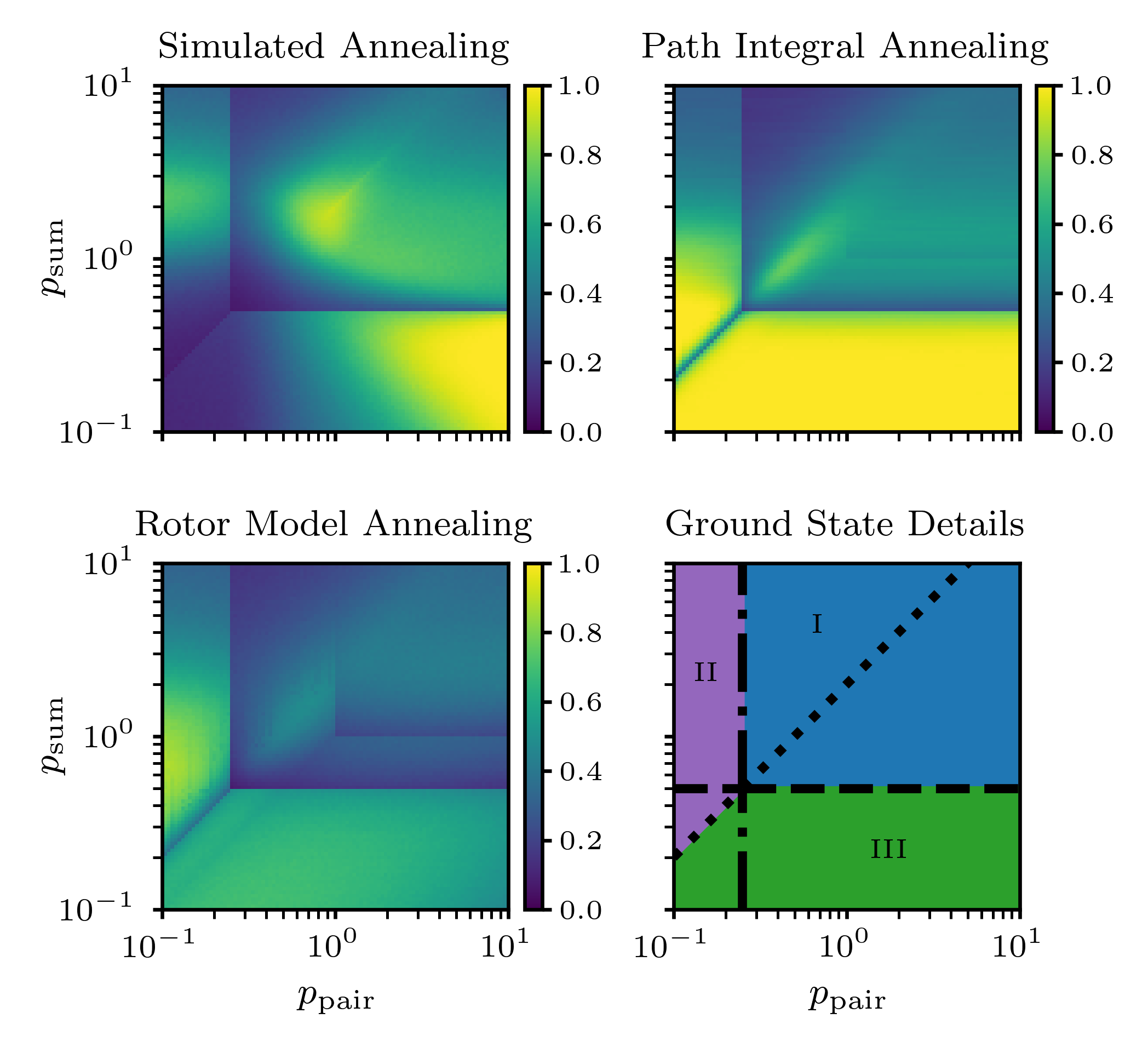}
    \caption{Probability that three different QUBO solvers provided by D-Wave find the optimal solution to the QUBO built from the 6 variable Job Shop instance. As the penalty values change, the identity of optimal solution to the QUBO changes. This is illustrated in the bottom right panel, with region I having a single optimal solution which is also feasible, and regions II and III having two or three degenerate infeasible solutions respectively. The black lines correspond to degeneracies between these sets of solutions.}
    \label{fig:ClassicalSolvers6QB}
\end{figure}

We expect that the different regions of the parameter space for a given problem instance will cause solvers to exhibit different behavior, though precisely what differences arise and how important they are will certainly depend on the solver implementation.
The most straightforward test of this expectation with any given solver is to perform a two-dimensional parameter sweep across $p_{\rm sum}$ and $p_{\rm pair}$ and plot the solution accuracy at each point.

We performed this test using three classical QUBO solvers provided by the D-Wave Ocean SDK~\cite{dwavesolvers} with their default settings, one based on simulated annealing and two based on simulated quantum annealing (referred to as ``path integral annealing'' and ``rotor model annealing'').
These three specific solvers were chosen for two reasons: firstly, they are all inspired by the physical annealing process and so will provide an interesting comparison for the results we obtain with quantum annealing; secondly, despite the simplicity and small size of the problem instances we tested with they do not always obtain the optimal solution and hence the solution accuracy is an easily accessed performance metric.
More complicated or better performing solvers could be used, but a more complicated metric would need to be found which effectively functions as a proxy for ``problem difficulty.''

The probability that each solver returns the optimal solution to the QUBO for a range of penalty parameters given $10^5$ trials is shown in Fig.~\ref{fig:ClassicalSolvers6QB}.
A variety of interesting behaviors are exhibited by the different solvers, but most important to note is that while they behave differently, all three solvers demarcate the transition between the QUBO solution being feasible and infeasible with a sharp transition in solution accuracy.
This is strong evidence that there is in fact some kind of transition in the difficulty of the problem at these points, and so we expect that similar features may be visible in the behavior of real, physical annealers such as quantum annealers like the D-Wave machines or CMOS annealers~~\cite{Chowdhury2025,Aadit22}.

\section{Thermodynamics of Quantum Annealing}
\label{sec:ThermodynamicsOfAnnealing}

\subsection{Experimental estimation}
The dynamics of the D\mbox{-}Wave quantum annealer are well approximated by a transverse--field Ising Hamiltonian given by 
\begin{align}
  H(s)
  &= A(s) H_{\rm mix} + B(s) H_{\rm cost} \nonumber
  \\
  &=
  - A(s)\sum_{i} \sigma_i^{x}
  + B(s)\left(
    \sum_{i} h_i \sigma_i^{z}
    +
    \sum_{\langle i,j\rangle} J_{ij}\, \sigma_i^{z}\sigma_j^{z}
  \right),
  \label{eq:dwave-hamiltonian}
\end{align}
where:
\begin{itemize}
  \item \(s \in [0,1]\) is the dimensionless annealing parameter controlled by the user via an annealing schedule.
  \item \(A(s)\) is the transverse--field energy scale, and \(B(s)\) is the problem (Ising) energy scale.
  \item \(\sigma_i^{x,z}\) are Pauli operators acting on qubit \(i\) with eigenvalues \(\pm 1\).
  \item \(h_i\) are programmable local fields and \(J_{ij}\) are programmable Ising couplings on the hardware graph.
\end{itemize}
We work in energy units such that the overall scale is dimensionless; in particular, we identify
\(
A(s) = \Gamma[1-s]
\)
and
\(
B(s) = s
\)
for some constant \(\Gamma\),
and in what follows we set \(\Gamma = 1\) for simplicity.

The control parameter \(s\) is steered as a function of physical time \(t\in[0,\tau]\) via an annealing schedule
\(s(t)\). D\mbox{-}Wave's \emph{forward} annealing corresponds to a monotonic ramp from the transverse--field
dominated regime to the problem Hamiltonian,
\begin{equation}
  s_{\mathrm{F}}(t) = \frac{t}{\tau},
  \qquad 0 \le t \le \tau.
\end{equation}
In this work we exploit \emph{reverse annealing}, which allows us to initialise the device in a chosen classical
spin configuration at \(s=1\), reduce the transverse field by decreasing \(s\) to a minimum value \(\bar{s}\), and then
return to \(s=1\). The reverse--anneal schedule used in the experiments is a symmetric, piecewise linear protocol
\begin{equation}
  s(t) =
  \begin{cases}
    1 - 2(1-\bar{s})\,\dfrac{t}{\tau}, & 0 \le t \le \dfrac{\tau}{2},\\[1.0em]
    -1 + 2\bar{s} + 2(1-\bar{s})\,\dfrac{t}{\tau}, & \dfrac{\tau}{2} \le t \le \tau,
  \end{cases}
  \label{eq:reverse-anneal-schedule}
\end{equation}
so that \(s(0) = s(\tau) = 1\) and \(s(\tau/2)=\bar{s}\).

We model the processor as initially prepared in a classical Gibbs state at inverse temperature \(\beta_1\) with
respect to the problem Hamiltonian at \(s=1\).
Concretely, for a classical spin configuration \(\{\sigma_i^z\}\), the corresponding Ising energy is
\begin{equation}
  E_z(\{\sigma_i^z\})
  =
  \sum_{i} h_i \sigma_i^z
  +
  \sum_{\langle i,j\rangle} J_{ij}\, \sigma_i^z\sigma_j^z.
\end{equation}
A thermal sample at inverse temperature \(\beta_1\) is generated by drawing configurations with probability
\begin{equation}
  p_{\beta_1}(\{\sigma_i^z\})
  =
  \frac{e^{-\beta_1 E_z(\{\sigma_i^z\})}}{Z(\beta_1)},
  \qquad
  Z(\beta_1) = \sum_{\{\sigma_i^z\}} e^{-\beta_1 E_z(\{\sigma_i^z\})}.
\end{equation}
On the hardware, the initial configuration is imposed at \(s=1\) using D\mbox{-}Wave's reverse annealing API
(``initial\_state" plus a custom ``anneal\_schedule"), so that the system begins in a classical state with energy
\(E_{1,\mathrm{i}} = E_z(\{\sigma_i^z\})\).

For a given choice of schedule \(s(t)\), minimum parameter \(\bar{s}\) (for reverse annealing), and annealing time
\(\tau\), we perform the following steps:
\begin{enumerate}
  \item Generate a set of initial spin configurations
        \(\{\sigma_i^z\}^{(k)}\) from the thermal distribution at \(\beta_1\).
  \item For each configuration, program the couplings \(\{h_i,J_{ij}\}\) and perform the chosen annealing schedule on the D\mbox{-}Wave quantum annealer.
  \item At the final time \(t=\tau\), read out the classical spin configuration
        \(\{\sigma_i^z\}_{\mathrm{f}}^{(k)}\) and compute the final energy
        \(E_{1,\mathrm{f}}^{(k)} = E_z(\{\sigma_i^z\}_{\mathrm{f}}^{(k)})\).
  \item Define the stochastic energy change of the processor as
        \begin{equation}
          \Delta E_1^{(k)} = E_{1,\mathrm{f}}^{(k)} - E_{1,\mathrm{i}}^{(k)}.
        \end{equation}
\end{enumerate}
For each parameter set we repeat this procedure over many independent runs to build up statistics of
\(\Delta E_1\). In the relaxation experiment (\(s(t)\equiv 1\)) we use \(1000\) independent initial configurations
and perform \(10\) anneals per configuration, yielding \(10^4\) samples for each value of \(\tau\).
A similar sampling strategy is used for the reverse annealing experiments as a function of \(\bar{s}\).

We model the processor (system 1) as an open quantum system weakly coupled to a thermal environment
(system 2) at temperature \(T_2\). The joint system--environment compound is assumed to be initially in a
factorised Gibbs state
\begin{equation}
  \rho_0
  =
  \frac{e^{-\beta_1 H_1}}{Z_1}
  \otimes
  \frac{e^{-\beta_2 H_2}}{Z_2},
\end{equation}
where \(H_1\) is the processor Hamiltonian at \(t=0\) (corresponding to \(s=1\)), \(H_2\) is the environment
Hamiltonian, \(\beta_i = 1/(k_B T_i)\) are the inverse temperatures, and \(Z_i = \mathrm{Tr}[e^{-\beta_i H_i}]\)
are the partition functions.

For a cyclic schedule such as the reverse anneal Eq.~\eqref{eq:reverse-anneal-schedule}, the joint statistics of the
energy changes \(\Delta E_1\) of the processor and \(\Delta E_2\) of the environment obey the multivariate
fluctuation theorem
\begin{equation}
  \frac{p(\Delta E_1,\Delta E_2)}{p(-\Delta E_1,-\Delta E_2)}
  =
  \exp\!\bigl(\beta_1 \Delta E_1 + \beta_2 \Delta E_2\bigr).
  \label{eq:multi-fluctuation}
\end{equation}
This is understood with respect to a two-point projective energy measurement scheme at the beginning and end
of the protocol.

Defining the stochastic total entropy production as
\begin{equation}
  \Sigma = \beta_1 \Delta E_1 + \beta_2 \Delta E_2,
\end{equation}
we obtain from \eqref{eq:multi-fluctuation} the second-law inequality
\begin{equation}
  \langle \Sigma \rangle = \beta_1 \langle \Delta E_1 \rangle + \beta_2 \langle \Delta E_2 \rangle \ge 0.
\end{equation}
Identifying the average heat absorbed by the processor from the environment with
\(\langle Q \rangle = - \langle \Delta E_2 \rangle\), and the average work performed on the processor + bath compound
by the external control with
\begin{equation}
  \langle W \rangle = \langle \Delta E_1 \rangle + \langle \Delta E_2 \rangle,
\end{equation}
one can classify the operation mode (refrigerator, engine, accelerator, heater) from the signs of
\(\langle \Delta E_1\rangle, \langle Q\rangle\) and \(\langle W\rangle\).

In practice, the environment energy change \(\Delta E_2\) is not directly measurable on the D\mbox{-}Wave device,
and the only experimentally accessible thermodynamic observable is the processor energy change \(\Delta E_1\).
To extract information about entropy production, heat, and work from \(\Delta E_1\) alone, we exploit a
thermodynamic uncertainty relation (TUR). Consider a joint distribution \(p(\sigma,\phi)\) satisfying
\begin{equation}
  \frac{p(\sigma,\phi)}{p(-\sigma,-\phi)} = e^{\sigma},
\end{equation}
then the TUR implies the bound
\begin{equation}
  \langle \sigma \rangle
  \ge
  2\, g\!\left(
    \frac{\langle \phi \rangle}{\sqrt{\langle \phi^2 \rangle}}
  \right),
  \qquad
  g(x) = x\,\tanh^{-1}(x).
\end{equation}
By choosing \(\sigma = \Sigma\) and \(\phi = \Delta E_1\) we obtain the lower bound on the average entropy production
\begin{equation}
  \langle \Sigma \rangle
  \ge
  2\, g\!\left(
    \frac{\langle \Delta E_1 \rangle}{\sqrt{\langle \Delta E_1^2 \rangle}}
  \right).
  \label{eq:entropy-bound}
\end{equation}
Combining \eqref{eq:entropy-bound} with the definitions of \(\langle Q \rangle\) and \(\langle W \rangle\) yields bounds
on the average heat and work
\begin{align}
  -\langle Q \rangle
  &\ge
  \frac{2}{\beta_2}
  g\!\left(
    \frac{\langle \Delta E_1 \rangle}{\sqrt{\mathrm{var}(\Delta E_1)}}
  \right) -\frac{\beta_1}{\beta_2} \langle \Delta E_1 \rangle,
  \label{eq:heat-bound}
  \\
  \langle W \rangle
  &\ge
  \frac{2}{\beta_2}
  g\!\left(
    \frac{\langle \Delta E_1 \rangle}{\sqrt{\mathrm{var}(\Delta E_1)}}
  \right)
  + \left(1 - \frac{\beta_1}{\beta_2}\right)\langle \Delta E_1 \rangle,
  \label{eq:work-bound}
\end{align}
where \(\mathrm{var}(\Delta E_1) = \langle \Delta E_1^{2} \rangle - \langle \Delta E_1 \rangle^{2}\). The thermodynamic efficency follows as:
\begin{equation}
    \eta \leq -\frac{ \langle W \rangle}{ \langle Q \rangle}.
    \label{efficiency}
\end{equation}

Thus, by measuring only the first two moments of \(\Delta E_1\) for each reverse annealing schedule,
we obtain lower bounds on the average entropy production, heat exchanged with the environment,
and work performed by the driving.

The thermodynamic bounds \eqref{eq:entropy-bound}--\eqref{eq:work-bound} depend on the bath inverse temperature
\(\beta_2\), which is not directly provided by the hardware. We estimate \(\beta_2\) from the final state statistics
by treating the measured spin configurations as approximate equilibrium samples of the classical Ising model at some effective temperature.

Let \(D = \{ \mathbf{s}^{(1)},\dots,\mathbf{s}^{(D)} \}\) be a dataset of \(D\) spin configurations obtained from the
device at \(s=1\), where \(\mathbf{s}^{(d)} = (s_1^{(d)},\dots,s_N^{(d)})\) with \(s_i^{(d)} \in \{-1,+1\}\).
Following the pseudo-likelihood approach for Ising models, we define the average pseudo-log-likelihood
\begin{align}
  &\Lambda(\beta)
  =
    \label{eq:pseudo-likelihood}
  \\ \nonumber
  &-\frac{1}{N D}
  \sum_{i=1}^{N}
  \sum_{d=1}^{D}
  \ln\!\left[
    1 + \exp\!\left(
      -2 \beta\, s_i^{(d)}\,
      \Bigl(
        h_i + \sum_{j \in \delta_i} J_{ij}\, s_j^{(d)}
      \Bigr)
    \right)
  \right],
\end{align}
where \(\delta_i\) denotes the set of neighbours of spin \(i\) in the problem graph.
The pseudo-likelihood estimator of the inverse temperature is then
\begin{equation}
  \hat{\beta}_2
  =
  \arg\max_{\beta} \Lambda(\beta).
  \label{eq:beta-estimator}
\end{equation}

Operationally, we proceed as follows:
\begin{enumerate}
  \item For a given minimum annealing parameter \(\bar{s}\), run the reverse annealing protocol and collect a large
        sample of final spin configurations at \(s=1\).
  \item Compute \(\Lambda(\beta)\) from \eqref{eq:pseudo-likelihood} for a range of \(\beta\) and find the maximiser
        \(\hat{\beta}_2\) according to \eqref{eq:beta-estimator}.
  \item Repeat this estimation for several values of \(\bar{s}\). For \(\bar{s} \lesssim 0.5\) the inferred \(\hat{\beta}_2\)
        is approximately independent of \(\bar{s}\); we therefore take the corresponding average value as our estimate
        of the environment inverse temperature.
\end{enumerate}

\subsection{Simulation and Numerical Estimation}

To simulate the quantum annealing process, we employed the adiabatic master equation as developed in Refs.~\cite{Albash2012}, with the annealing Hamiltonian as given in Eq.~\eqref{eq:dwave-hamiltonian}.
For forward annealing the annealing parameter $s$ increases linearly as $t/\tau$ with $\tau$ being the annealing time, while for reverse annealing $s$ decreases linearly from 1 to some setpoint $0<s'<1$ then returns to 1 after a configurable pause duration.
In our simulations, the Hamiltonian coefficients $A(s), B(s)$ are interpolated from the table provided by D-Wave describing the annealing schedule implemented on their hardware~\cite{dwaveannealingschedule}.

In simulation, we have complete control over the initial state input into the annealing process. 
We use thermal states at a configurable inverse temperature $\beta$, for forward annealing defined using the mixer Hamiltonian $H_{\rm mix}$ and for reverse annealing defined using the cost Hamiltonian $H_{\rm cost}$. 

Once the initial state has been chosen and the annealing schedule finalized, the simulation proceeds by integrating the master equation over a series of small fixed intervals $\delta t = \tau/N$ using tight numerical tolerances to compute a series of density matrices $\rho(t_n)$ giving the state of the system throughout the annealing process.
This is a noisy simulation, so the annealing process is simulated assuming that the qubits are in contact with some bath.
Following Ref.~\cite{Albash2012}, the qubits are weakly-coupled to an Ohmic bath with an interaction strength of $10^{-4}$, a cutoff frequency of $\SI{4}{\giga\hertz}$ at a temperature of $\SI{16}{\milli\kelvin}$, which is reasonable for comparison to D-Wave's quantum annealers~\cite{Albash2015,Albash2015b}.
This time series is then post-processed to reconstruct the thermodynamic quantities in which we are interested, both directly and using the same thermodynamic uncertainty relation-based approach as employed when analyzing our experimental data.

\subsubsection{Post-processing}

Computing bounds on the entropy production, heat flux, and work using the thermodynamic uncertainty relations from our simulation results is straightforward. 
The initial temperature $\beta$ and environment temperature $\beta_{\rm env}$ are known, so all that needs to be computed are the expected first and second moments of the energy change, $\ev{\Delta E}$ and $\ev{\Delta E^2}$. 
Given the initial and final states $\rho(0)$ and $\rho(\tau)$ along with the initial and final Hamiltonians $H(0)$ and $H(\tau)$, these values are computed as 
\begin{subequations}
\begin{align}
    \ev{\Delta E} &= \tr\bqty*{H(\tau)\rho(\tau)} - \tr\bqty{H(0)\rho(0)}
    ,
    \\
    \ev{\Delta E^2} &= \tr\bqty*{H(\tau)^2\rho(\tau)} + \tr\bqty*{H(0)^2\rho(0)} \\
            \nonumber &\qquad- 2\tr\bqty*{H(\tau)\rho(\tau)}\tr\bqty*{H(0)\rho(0)}
    .
\end{align}
\end{subequations}

A more detailed analysis of the simulation output allows the change in entropy, heat, and work across each time step to be estimated directly.
The easiest of these three quantities to estimate is the entropy production, which we calculate by simply computing the change entropy between each pair of simulated output:
\begin{align}
    \Sigma(t_k)\delta t \approx -\tr\bqty{\rho(t_{k+1})\log\rho(t_{k+1})} + \tr\bqty{\rho(t_k)\log\rho(t_k)}
    .
\end{align}
\begin{figure*}
    \centering
    \includegraphics[width=\linewidth]{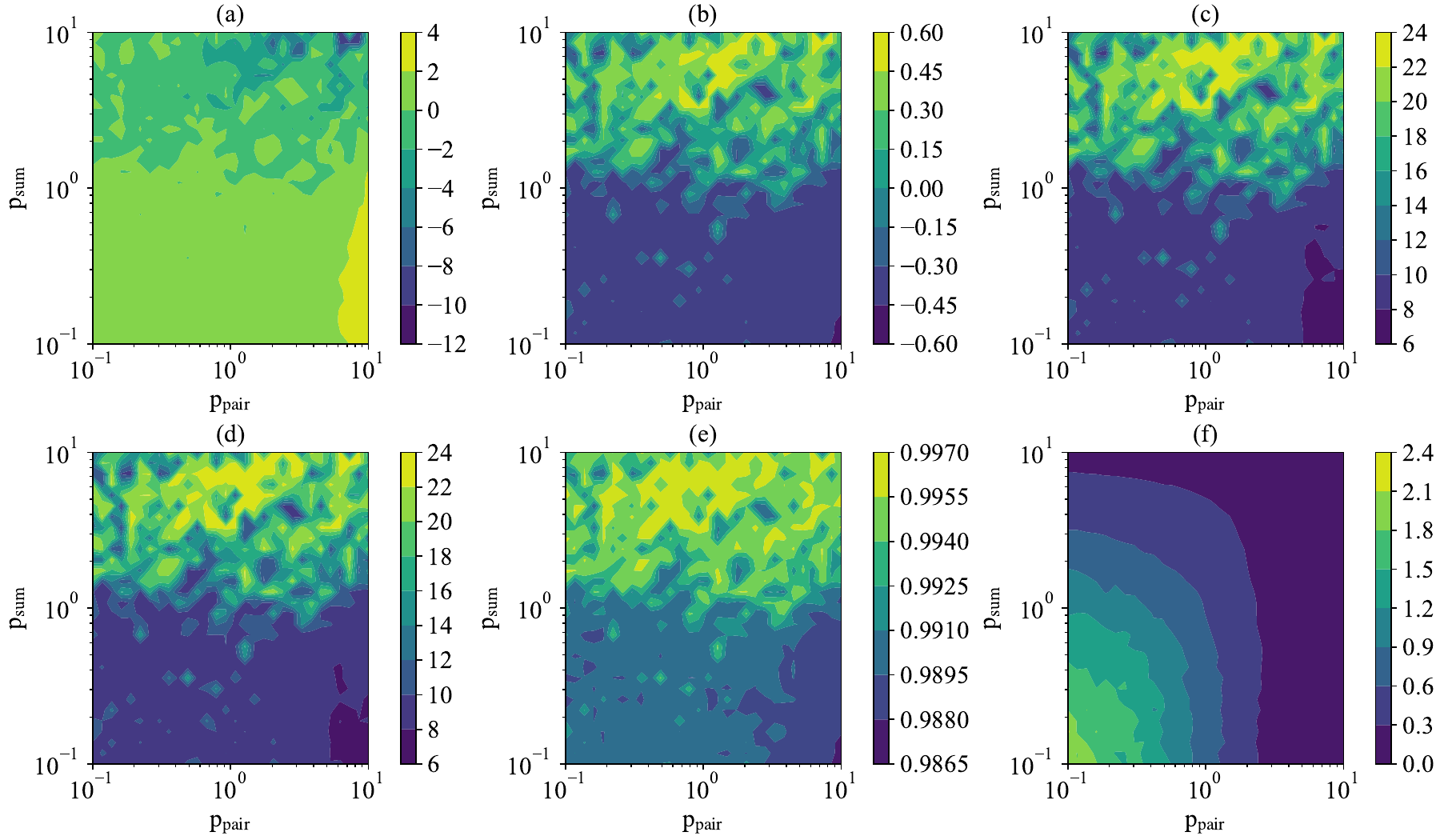}
    \caption{\textbf{Thermodynamics of the Job Shop problem on D-Wave Advantage before the critical region ($\bar s=0.15$).}
Job Shop (10 qubits) executed via reverse annealing on the D-Wave Advantage system. The processor is initialized in a classical Gibbs state at inverse temperature $\beta_1=10$ and reverse annealed for $t_a=10~\mu$s down to $\bar s=0.15$ (before the quantum critical region) and back to $s=1$.
Each panel is plotted versus the QUBO penalty weights $(p_{\rm pair},p_{\rm sum})$ controlling precedence and one-hot constraints.
(a) mean processor energy change $\langle\Delta E_1\rangle=\langle E_{1,f}-E_{1,i}\rangle$;
(b) TUR-based lower bound on the average entropy production~\eqref{eq:entropy-bound};
(c) work bound~\eqref{eq:work-bound};
(d) heat bound~\eqref{eq:heat-bound};
(e) thermodynamic efficiency~\eqref{efficiency};
(f) effective bath inverse temperature $\hat\beta_2$ obtained from pseudo-likelihood estimation~\eqref{eq:beta-estimator}.
The maps reveal a pronounced regime boundary driven primarily by $p_{\rm sum}$: strengthening constraint penalties reorganizes the energy landscape and produces a concomitant change in dissipation (entropy/work/heat) and efficiency, directly demonstrating that QUBO encoding parameters act as thermodynamic control knobs on hardware.}
    \label{fig4}
\end{figure*}
The heat and work flux in each timestep are somewhat more complex to compute. 
Given the initial and final states of any timestep $\rho(t_k)$ and $\rho(t_{k+1})$, we can easily compute the change in energy across this interval as,
\begin{align}
    \Delta E(t_{k+1}, t_k) = \tr\bqty{H(t_{k+1})\rho(t_{k+1})} - \tr\bqty{H(t_k)\rho(t_k)}
    ,
\end{align}
however this is the total change in energy which incorporates both heat and work which we must disentangle.
We can always write a quantum map which represents the time evolution across this timestep,
\begin{align}
    \rho(t_{k+1}) = \mathcal{E}_{t_{k+1}, t_k}\bqty{\rho(t_{k})}
    ,
\end{align}
which we approximate as being built from the composition of two separate quantum maps,
\begin{align}
    \rho(t_{k+1}) \approx \mathcal{V}_{t_{k+1},t_k}\circ\mathcal{U}_{t_{k+1},t_k}\bqty{\rho(t_{k})}
    .
\end{align}
The quantum map $\mathcal{U}$ represents the unitary evolution of the system under the time-dependent Hamiltonian,
\begin{align}
    \mathcal{U}_{t_{k+1},t_k}\bqty{\rho(t_{k})} &= U(t_{k+1},t_k) \rho(t_{k}) U^\dagger(t_{k+1},t_k)
    ,
\end{align}
with
\begin{align}
    U(t_{k+1},t_k) &= \exp\bqty{-i\int_{t_k}^{t_{k+1}} \dd{t'} H(t')}
    ,
\end{align}
and the noise contribution is collected into the quantum map $\mathcal{V}$ corresponding to the noise terms only of the adiabatic master equation~\cite{Albash2012}.
Assuming the timestep $\delta t$ is sufficiently small, the error in this decomposition is small and so we may analyze the unitary and non-unitary evolutions separately in a similar fashion to some analyses of fluctuation theorems~\cite{Crooks1998,Jarzynski1997}.
The work corresponds to the energetic cost of changing the annealing parameter, which we identify with the change in energy due only to the unitary evolution,
\begin{align}
    W(t_{k+1},t_k) = \tr\bqty{H(t_{k+1}) \mathcal{U}_{t_{k+1},t_k}\bqty{\rho(t_{k})}} - \tr\bqty{H(t_k)\rho(t_k)}
    ,
\end{align}
where $\mathcal{U}_{t_{k+1},t_k}\bqty*{\rho(t_k)}$ can be computed by simulating the time-dependent Schr{\"o}dinger equation.
Finally, knowing the total change in energy and the work performed in this timestep, the heat exchanged with the environment is their difference according to the first law,
\begin{align}
    Q(t_{k+1},t_k) = \Delta E(t_{k+1},t_{k}) - W(t_{k+1}, t_k)
    .
\end{align}

\section{Results}
\label{sec:Results}

\subsection{Experimental results}
\begin{figure*}
    \centering
    \includegraphics[width=\linewidth]{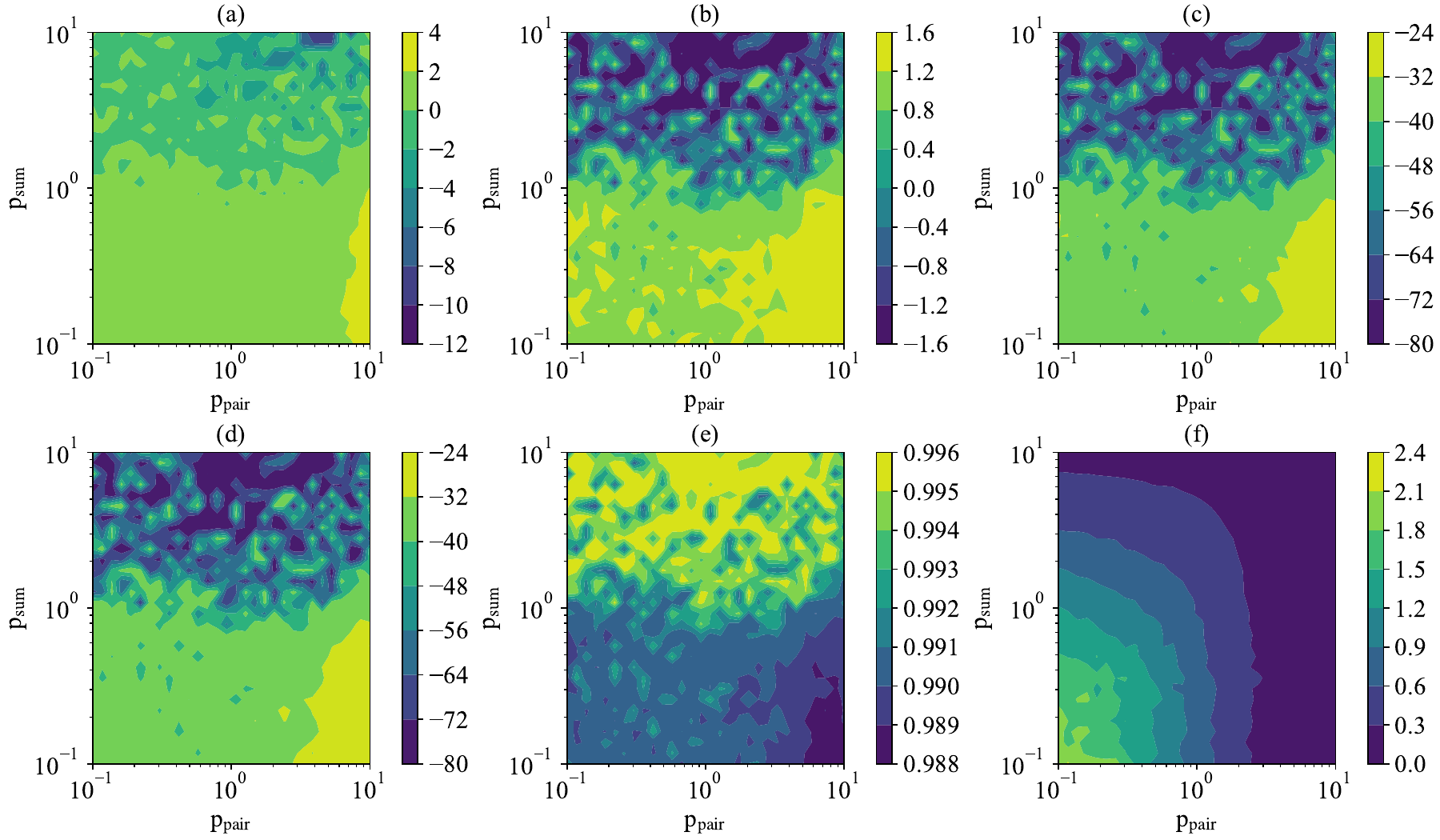}
    \caption{\textbf{Thermodynamics of the Job Shop problem on D-Wave Advantage near the critical region ($\bar s=0.27$).}
Job Shop (10 qubits) executed via reverse annealing on the D-Wave Advantage system. The processor is initialized in a classical Gibbs state at inverse temperature $\beta_1=10$ and reverse annealed for $t_a=10~\mu$s down to $\bar s=0.27$ (near the quantum critical region) and back to $s=1$.
Each panel is plotted versus the QUBO penalty weights $(p_{\rm pair},p_{\rm sum})$ controlling precedence and one-hot constraints.
(a) mean processor energy change $\langle\Delta E_1\rangle=\langle E_{1,f}-E_{1,i}\rangle$;
(b) TUR-based lower bound on the average entropy production~\eqref{eq:entropy-bound};
(c) work bound~\eqref{eq:work-bound};
(d) heat bound~\eqref{eq:heat-bound};
(e) thermodynamic efficiency~\eqref{efficiency};
(f) effective bath inverse temperature $\hat\beta_2$ obtained from pseudo-likelihood estimation~\eqref{eq:beta-estimator}.
Relative to $\bar s=0.15$, the entropy/work/heat landscapes display enhanced spatial variability and sharper local features, consistent with increased susceptibility to thermal noise and decoherence near small-gap dynamics.
Nevertheless, the same $p_{\rm sum}$-dominated regime structure persists, indicating that encoding-induced spectral transitions remain visible thermodynamically even in the most noise-sensitive operating region.}
    \label{fig5}
\end{figure*}
\begin{figure*}
    \centering
    \includegraphics[width=\linewidth]{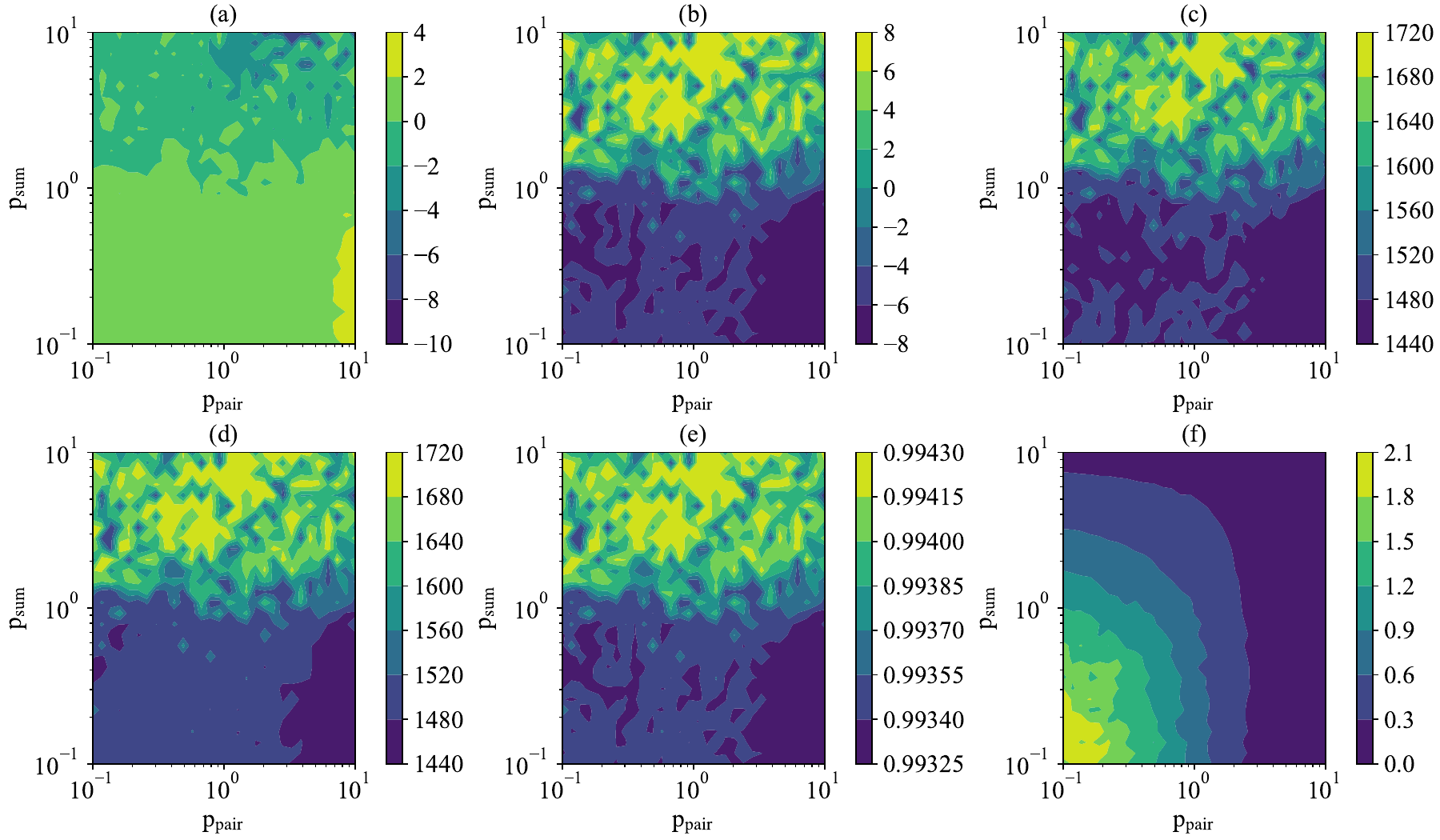}
    \caption{\textbf{Thermodynamics of the Job Shop problem on D-Wave Advantage after the critical region ($\bar s=0.35$).}
Job Shop (10 qubits) executed via reverse annealing on the D-Wave Advantage system. The processor is initialized in a classical Gibbs state at inverse temperature $\beta_1=10$ and reverse annealed for $t_a=10~\mu$s down to $\bar s=0.35$ (after the quantum critical region) and back to $s=1$.
Each panel is plotted versus the QUBO penalty weights $(p_{\rm pair},p_{\rm sum})$ controlling precedence and one-hot constraints.
(a) mean processor energy change $\langle\Delta E_1\rangle=\langle E_{1,f}-E_{1,i}\rangle$;
(b) TUR-based lower bound on the average entropy production~\eqref{eq:entropy-bound};
(c) work bound~\eqref{eq:work-bound};
(d) heat bound~\eqref{eq:heat-bound};
(e) thermodynamic efficiency~\eqref{efficiency};
(f) effective bath inverse temperature $\hat\beta_2$ obtained from pseudo-likelihood estimation~\eqref{eq:beta-estimator}.
The efficiency becomes comparatively uniform across encoding space while the work/heat bounds can grow in magnitude, consistent with a more classical freeze-out dominated by dissipation set by the effective temperature $\hat\beta_2$.
Together with~\Cref{fig4,fig5}, these results show that both the encoding parameters $(p_{\rm sum},p_{\rm pair})$ and the dynamical operating point $\bar s$ determine the irreversibility and energetic cost of quantum annealing.}

    \label{fig6}
\end{figure*}
~\Cref{fig4,fig5,fig6} report the experimentally inferred thermodynamic observables for the 10-qubit Job Shop instance implemented on the D-Wave Advantage processor under reverse annealing. For each encoding point $(p_{\rm sum},p_{\rm pair})$, we prepare initial classical configurations from a Gibbs distribution at inverse temperature $\beta_1$ with respect to the programmed Ising cost Hamiltonian at $s=1$, execute the cyclic reverse-anneal schedule down to a minimum $\bar s$ and back, and record the stochastic processor energy change $\Delta E_1 = E_{1,f}-E_{1,i}$. From the first two moments of $\Delta E_1$ we compute the TUR-based lower bound on the average entropy production~\eqref{eq:entropy-bound} and the corresponding bounds on work~\eqref{eq:work-bound} and heat~\eqref{eq:heat-bound}, using an effective bath inverse temperature $\hat\beta_2$ estimated from output samples via the pseudo-likelihood procedure~\eqref{eq:beta-estimator}. While these quantities are bounds (rather than direct calorimetric measurements), their \emph{dependence} on the encoding parameters provides an operational probe of how the QUBO mapping reshapes dissipation and irreversibility in hardware.

Across all three minimum annealing points $\bar s$ (before, near, and after the device's quantum critical region), the maps in~\Cref{fig4,fig5,fig6} exhibit a pronounced separation between a low-$p_{\rm sum}$ regime and a higher-$p_{\rm sum}$ regime. This is the thermodynamic counterpart of the ``feasible vs.\ infeasible'' and ``split vs.\ unsplit'' boundaries diagnosed spectrally and algorithmically in ~\Cref{fig:Splitness6QB,fig:ClassicalSolvers6QB}: when constraint penalties are too weak, low-energy configurations that violate one-hot/sum constraints proliferate and dominate the low-energy landscape, whereas sufficiently strong penalties lift the infeasible manifold and isolate the feasible optimum. In the thermodynamic data, this transition manifests not primarily as a change in a single scalar, but as a coordinated reorganization across $\langle\Delta E_1\rangle$ (panel a), the TUR entropy bound (panel b), and the work/heat bounds (panels c,d), indicating that the \emph{encoding-controlled spectral rearrangement} simultaneously changes (i) which states are visited and (ii) how strongly the dynamics couple to the environment.

The observed asymmetry between $p_{\rm sum}$ and $p_{\rm pair}$ is expected on structural grounds: the $p_{\rm sum}$ penalty contributes negative diagonal terms absent from the pairwise-penalty structure, so too-small $p_{\rm sum}$ can be ``overruled'' by objective contributions, yielding a qualitatively altered spectrum with increased degeneracy and mixed feasible/infeasible ordering. Thermodynamically, increased degeneracy near the bottom of the spectrum enhances the density of near-resonant transitions and provides more channels for bath-induced transitions during the protocol, which raises irreversibility and reshapes the balance between work-like driving costs and heat exchange. This explains why the most visible boundaries in the thermodynamic observables run predominantly along the $p_{\rm sum}$ axis, whereas variation in $p_{\rm pair}$ mainly modulates these features rather than setting them.

Panel (a) measures whether the processor typically returns to $s=1$ at a lower or higher Ising energy than it started with. In the weak-penalty regime, $\langle\Delta E_1\rangle$ is comparatively uniform, consistent with a landscape where constraint violations allow many low-barrier moves that do not systematically relax toward the feasible optimum. As $p_{\rm sum}$ increases into the feasible regime, $\langle\Delta E_1\rangle$ changes more strongly and becomes spatially structured, reflecting the emergence of a constrained manifold and a modified set of relaxation pathways. Importantly, the fine-grained ``speckling'' seen in the high-$p_{\rm sum}$ region is a hallmark of hardware sensitivity: once the feasible manifold is energetically isolated, small perturbations to programmed couplings (control noise, calibration drift, crosstalk) can reorder low-lying states and alter which basin the system freezes into, producing point-to-point variability that is absent in idealized simulations.

The TUR bound in panel (b) and the derived work and heat bounds in panels (c,d) show their strongest spatial variation near the same encoding crossover where solver accuracy exhibits sharp transitions in Fig.~\eqref{fig:ClassicalSolvers6QB}. This is physically consistent: at the crossover, objective and penalty terms compete on comparable energy scales, producing a denser set of avoided crossings and smaller effective gaps along the annealing path. In an open system, small gaps and dense spectra increase susceptibility to both thermal excitations and dephasing, which enhances entropy production and increases the energetic cost of driving. From the standpoint of the TUR estimator, this is compounded by the fact that the bound is controlled by the ratio of the mean to the variance of $\Delta E_1$; near a crossover, $\mathrm{var}(\Delta E_1)$ typically grows due to heterogeneous freeze-out pathways and sample-to-sample variability, which directly loosens the bound and accentuates apparent thermodynamic ``hot spots''.

Panel (e) reports the thermodynamic efficiency~\eqref{efficiency} computed from the work and heat estimates. The salient point is not the absolute proximity to unity—which is expected given that these are bound-based quantities in dimensionless energy units—but rather the \emph{relative ordering} across encoding space: the efficiency is systematically reduced in the same regions where entropy production and work/heat costs are enhanced. This anticorrelation supports the central thesis of the paper: encoding choices that make the optimization ``hard'' also make the annealing cycle more irreversible and therefore less thermodynamically efficient.

The inferred $\hat\beta_2$, Eq.\eqref{eq:beta-estimator}, exhibits a comparatively smooth dependence on $(p_{\rm sum},p_{\rm pair})$ and is largely insensitive to $\bar s$ for $\bar s \lesssim 0.5$, consistent with the empirical procedure used to estimate it. A key interpretation is that $\hat\beta_2$ should be regarded as an \emph{effective} temperature that absorbs multiple non-idealities: not only the physical fridge temperature, but also analog control errors and the encoding-dependent rescaling required to fit coefficients into the hardware range. In particular, increasing penalty magnitudes typically forces stronger global rescaling of all couplings, reducing the physical energy scale of the problem Hamiltonian relative to thermal fluctuations and low-frequency noise; this appears as a reduced $\hat\beta_2$ (higher effective temperature). This mechanism provides a direct thermodynamic route by which ``over-penalization'' can degrade both solution quality and efficiency: it can simultaneously isolate the feasible manifold in the ideal spectrum while making it less distinguishable under thermal/noisy sampling on hardware.

Comparing~\Cref{fig4,fig5,fig6} highlights how the same encoding interacts differently with the device depending on whether the reverse anneal penetrates (i) the transverse-field-dominated regime (Fig.~\eqref{fig4}, $\bar s=0.15$), (ii) the critical region where driver and problem terms compete most strongly (Fig.~~\eqref{fig5}, $\bar s=0.27$), or (iii) the more classical regime after the critical point (Fig.~~\eqref{fig6}, $\bar s=0.35$). Near the critical region the instantaneous gap is expected to be smallest and the system most sensitive to both thermal excitation and decoherence, which naturally explains why the most pronounced spatial fluctuations and sign changes in bound-based thermodynamic quantities appear at $\bar s=0.27$. For $\bar s=0.35$, the protocol spends less time in the strongly quantum regime and the dynamics are expected to freeze out in a more classical manifold; correspondingly, the efficiency landscape becomes more uniform even though the work/heat bounds can grow in magnitude (partly reflecting the $\hat\beta_2$ normalization in the bounds). Overall, these trends support an operational picture in which encoding controls the spectrum, while $\bar s$ controls which portion of that spectrum is dynamically relevant before freeze-out.

The clean boundaries and sharp transitions in~\Cref{fig:Splitness6QB,fig:ClassicalSolvers6QB} arise from idealized access to the encoded objective landscape (exact energies, exact penalties, and algorithmic solvers that do not suffer analog control errors). In contrast, the hardware maps in~\Cref{fig4,fig5,fig6} fold in several physical effects that systematically smear and shift these transitions:
(i) \emph{analog programming errors} ($h_i\!\to\! h_i+\delta h_i$, $J_{ij}\!\to\! J_{ij}+\delta J_{ij}$) that are effectively amplified when large penalty magnitudes enforce strong global rescaling;
(ii) \emph{open-system decoherence and thermalization}, which are most consequential near small-gap regions and can drive population away from the ideal ground state during the reverse cycle;
(iii) \emph{encoding-dependent freeze-out}, where the output distribution reflects the state at an earlier $s^\ast$ that depends on the local gap structure and relaxation times, rather than an equilibrium distribution at $s=1$;
and (iv) \emph{estimation bias in $\hat\beta_2$}, since the pseudo-likelihood procedure assumes approximate Gibbsian sampling, whereas the device outputs are generally non-equilibrium samples influenced by freeze-out and readout noise. These effects do not negate the regime structure predicted by the idealized maps; instead, they explain why the experimental thermodynamic signatures are broadened, more heterogeneous, and sometimes shifted in $(p_{\rm sum},p_{\rm pair})$ relative to the solver-based boundaries. Crucially, the persistence of a correlated crossover across energy change, entropy production, and efficiency demonstrates that the encoding-driven spectral transition remains visible even under realistic noise and decoherence.

\subsection{Simulation results}

\begin{figure}[t!]
    \centering
    \includegraphics[width=\columnwidth]{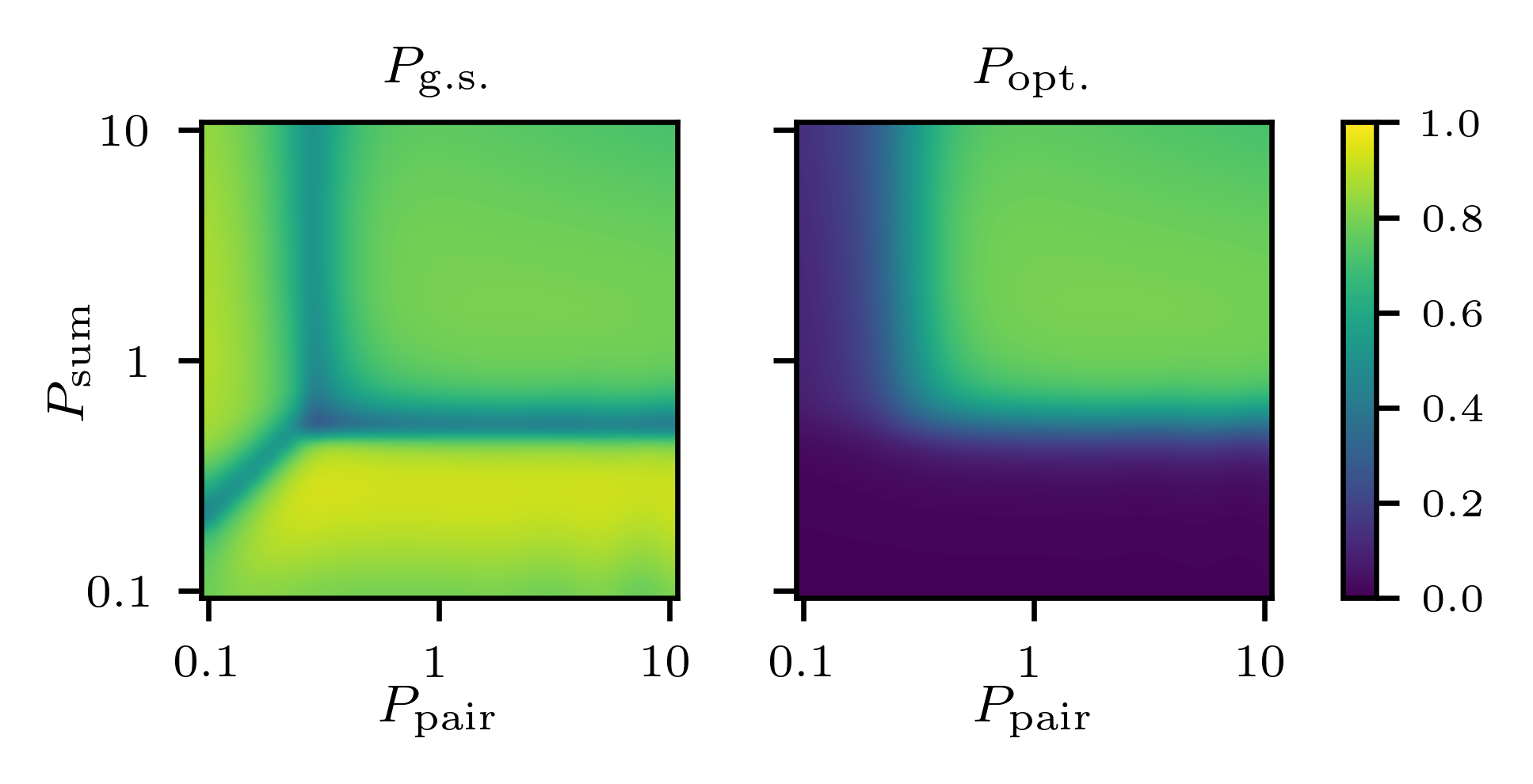}
    \caption{Probability of finding the (left) overall ground state or (right) best feasible solution for the 4 variable instance according to numerical simulations with an initial temperature of $\beta = 10$ and annealing time of $\tau = \SI{10}{\nano\second}$.}
    \label{fig:SimulatedProb4QB}
\end{figure}
\begin{figure}[t!]
    \centering
    \includegraphics[width=\columnwidth]{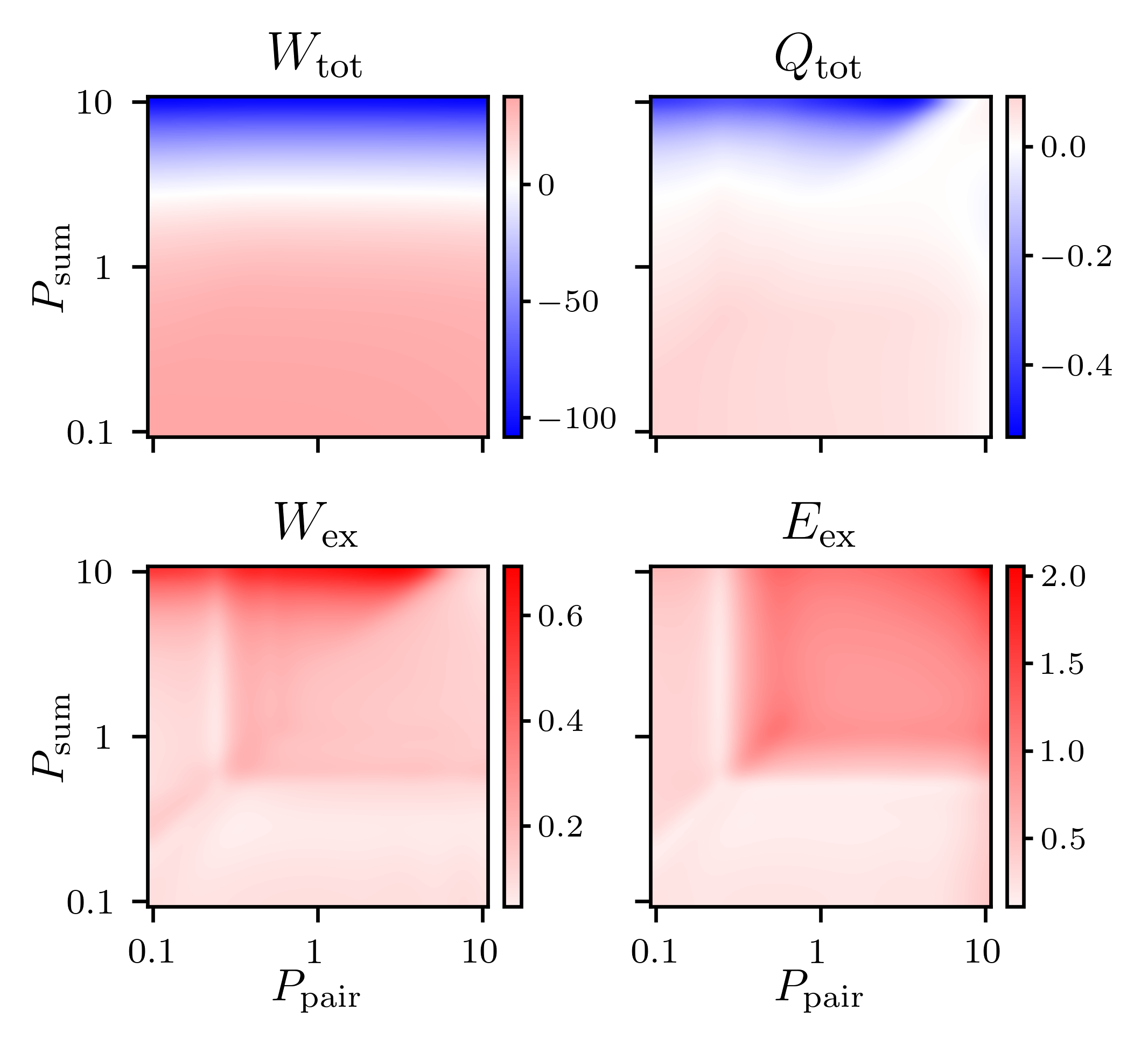}
    \caption{Numerical estimation of the total work and heat flux as well as the excess work and excess energy according to simulations of the annealing process with $\beta = 10$ and $\tau = \SI{10}{\nano\second}$.}
    \label{fig:SimulatedThermo4QBb}
\end{figure}

Figure~\ref{fig:SimulatedProb4QB} shows the probabilities of obtaining the ground state of the annealing problem as well as the probability of obtaining the optimal solution to the Job Shop problem. 
As in the case of the results obtained with the three classical annealing solvers provided by the D-Wave SDK shown in Fig.~\ref{fig:ClassicalSolvers6QB}, we find that quantum annealing also shows a sharp drop in the solution probability corresponding to the points where the ground state transitions from feasible to infeasible.
Physically, this is due to the gap between the ground state and lowest-lying excited states vanishing at these points.
The annealing time would need to be drastically increased near these transitions to find the solution accurately, however with longer annealing times and a corresponding increase in the effect of noise on the optimization process there will be a limit beyond which features wash out.

One important difference between the simulation results and the results of the three classical solvers tested earlier is that we do not observe much detail in the solution probability within the feasible region of the penalty parameter space with quantum annealing, while significant features are visible in the results of the classical solvers in Fig.~\ref{fig:ClassicalSolvers6QB} (note some differences are because the classical solvers were applied to a 6 qubit instance vs. the 4 qubit instance simulated for quantum annealing).
One explanation for this is that away from the boundary of the feasible region the spectral gap of the cost Hamiltonians generated from our Job Shop problems are constant.
The spectral gap plays a key role in determining the accuracy of the annealing process \cite{Aharonov2008,Lidar2009,Boixo2010}, hence we expect comparatively minor performance variations when the gap is fixed.
This is an important distinction between quantum annealing and the three classical annealing models tested, and may indicate that the performance of quantum annealing can generally be less sensitive to structural properties of the optimization problem to solve.

The thermodynamic quantities (total work, total heat, excess work, energy excess) computed from our simulations are shown in Fig.~\ref{fig:SimulatedThermo4QBb}.
Quantitatively there are important differences compared with our results obtained from tests with the real device presented earlier, but that should not be surprising as we simulated forward annealing of a much smaller problem than was tested on the real machine with reverse annealing.
Qualitatively, we observe the same general structure: for the overall annealing process, the sum penalty terms have a far more pronounced effect than the pair penalty terms. 
We do \emph{not} see strong features corresponding to transitions in the optimization problem as might be expected from the solution probabilities of Fig.~\ref{fig:SimulatedProb4QB}, however this agrees with the fact that we do not see such features in our experimental results. 
Importantly, this indicates that we can use results obtained with TUR-based estimates of thermodynamic quantities obtained with reverse annealing (the limit of what is possible with current-generation devices) to shed light on the overall thermodynamics of both the forward and reverse annealing processes and how they might depend on penalty and embedding parameters.

Additionally, when we look at more fine-grained thermodynamic quantities of excess work and energy, we now are able to see features corresponding to level crossings and transitions in the identity of the ground states of the Ising model.
In fact, we see more transitions than just those corresponding to the feasible-infeasible transition.
Faint features are visible which correspond to degeneracies and crossings of higher-lying levels and points where various subgaps close -- similar to the structures visible in the solution probabilities of the classical solvers from Fig.~\ref{fig:ClassicalSolvers6QB}.
This shows that while we do not see strong indications of detailed structure of the solution probability of the quantum annealer, such details do still affect the quantum annealing process and ``how difficult'' the problem is to solve.
The effect is simply much weaker, again hinting that quantum annealing may be less sensitive to such details as compared to classical annealing solvers.

\section{Conclusion}
\label{sec:Conclusion}

We established that QUBO encoding is not a neutral preprocessing step, but rather a physically consequential design choice that controls both (i) the computational hardness encountered by a quantum annealer and (ii) the irreversibility and energetic cost of the annealing dynamics when the device is viewed as an open thermodynamic machine. Since a single constrained optimization task admits many QUBO realizations, and because present-day annealers operate in a regime where finite temperature, control errors, and decoherence are non-negligible, different encodings of the ``same'' problem can lead to measurably different outcomes in both success probability and dissipative cost.

Focusing on a Job Shop Scheduling instance, we constructed an encoding family parameterized by penalty weights $p_{\rm sum}$ (one-hot / sum constraints) and $p_{\rm pair}$ (precedence constraints), and used these parameters as a controlled way to reshape the encoded Ising spectrum.
We showed that the $(p_{\rm sum},p_{\rm pair})$ plane naturally separates into distinct encoding regimes, including a feasibility transition where the QUBO ground state ceases to represent a valid schedule if penalties are too weak, and a related ``mixed'' regime where feasible and infeasible low-energy states are no longer energetically segregated. 

These regime boundaries are not merely algebraic artifacts: they coincide with sharp changes in solver performance and with qualitatively distinct thermodynamic signatures, supporting the central message that encoding parameters simultaneously determine computational hardness and thermodynamic efficiency. Furthermore, we extracted experimentally relevant thermodynamic information from limited hardware observables. On the device, the environment energy change is not directly accessible; instead, we measure only the processor energy change $\Delta E_1$ at the end of a cyclic reverse-annealing protocol. 
By leveraging a thermodynamic uncertainty relation, we obtain a lower bound on the mean entropy production and corresponding bounds on average work and heat in terms of the first two moments of $\Delta E_1$, together with an effective bath inverse temperature estimated from output statistics.
This provides a practical and device-agnostic framework for mapping encoding space into a thermodynamic phase diagram of dissipation and efficiency, directly enabling a ``thermodynamics-aware'' notion of encoding quality.

Our results also clarify why tuning constraint strengths is intrinsically an energy-scale matching problem in NISQ hardware. If penalties are too small, the unconstrained optimum can violate the original constraints; if penalties are too large, the objective becomes a small perturbation that can be washed out by noise and finite-temperature sampling.
Moreover, the encoding dependence is generically asymmetric in $p_{\rm sum}$ and $p_{\rm pair}$ because the sum-constraint structure introduces negative diagonal contributions that can be overruled by the objective when $p_{\rm sum}$ is too small, whereas the pairwise penalties do not modify the diagonal in the same way; this structural asymmetry provides a microscopic explanation for the stronger leverage of $p_{\rm sum}$ on both difficulty and thermodynamic cost. 

From the thermodynamic perspective, the optimal encoding is therefore not simply the one with ``largest penalties,'' but rather the one that \emph{balances} penalties to (i) enforce feasibility and energetic separation of constraint-violating configurations while (ii) maintaining an effective problem energy scale that remains resolvable against thermal fluctuations and analog errors, thereby minimizing dissipated work and entropy production. 

Beyond the specific Job Shop instance studied here, the broader implication is that QUBO design should be treated as a \emph{co-design problem} between algorithm and hardware. Any encoding reshapes degeneracies, barrier structure, and avoided-crossing patterns along the annealing path, and these spectral features govern not only the probability of reaching the desired ground state but also the extent to which the dynamics are driven out of equilibrium, generating heat and entropy.
This thermodynamic viewpoint suggests new and experimentally testable encoding principles: for example, one may prefer encodings that avoid dense low-energy manifolds and reduce the competition between objective and penalties in the regime where the annealer is most susceptible to environmental noise, thereby improving both solution quality and energetic efficiency.

Several important directions follow from this work. First, it will be valuable to extend the present ``encoding thermodynamics'' analysis to larger instances and to different constraint structures, to test which observed correlations persist with scaling and embedding overhead. Second, while TUR-based quantities provide robust bounds from minimal data, tightening the connection between bound-based efficiency metrics and operational measures (time-to-solution, hybrid post-processing cost, or hardware energy consumption) would sharpen their utility for practical encoding selection. Finally, improving open system models and calibration-aware simulations—especially in regimes where device sampling deviates from Gibbs behavior—will help disentangle intrinsic spectral effects from hardware-specific control noise and freeze-out, strengthening the predictive power of thermodynamics-inspired encoding optimization.

\acknowledgments{%
The authors acknowledge the \href{https://www.fz-juelich.de/ias/jsc}{J\"ulich Supercomputing Centre} for providing computing time on the D-Wave Advantage™ System JUPSI through the J\"ulich UNified Infrastructure for Quantum computing (JUNIQ). 

E.D. acknowledges U.S. NSF under Grant No. OSI-2328774. %
K.D. acknowledges: Scientific work co-financed from the state budget under the program of the Minister of Education
and Science, Poland (pl. Polska) under the name "Science for Society II" project number NdS-II/SP/0336/2024/01 funding amount 1000000 PLN  
total value of the project 1000000 PLN  \raisebox{-4pt}{\includegraphics[width = 0.15\textwidth]{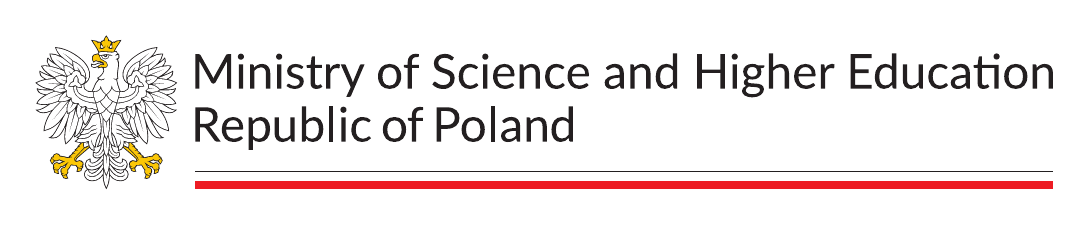}}
K.D. acknowledges the consultation with the railway operator \href{https://www.kolejeslaskie.pl/en/}{Koleje Śląskie sp. z o.o.}, on practical aspects of disturbance mitigation. %
B.G. acknowledges the Sonata Bis 10 project, No. 2020/38/E/ST3/00269. %
Z.M. acknowledges funding from the Ministry of Economic Affairs, Labour and Tourism Baden-Württemberg in the frame of the Competence Center Quantum Computing Baden-Württemberg (project ``KQCBW25'').

GitHub Link: \url{https://github.com/iitis/QUBO-encoding-effects-on-quantum-annealing-efficiency}
}

\appendix

\section{ILP Encoding}
\label{app:ILPEncoding}

Following the standard encoding of such problems~\cite{pinedo2012scheduling} we use the following definitions:
\begin{itemize}
\item Let $S_j$ be the schedule of job $j$, i.e. the series of machines it must be processed on.
\item Let $p_{j,m}$ be the processing time of job $j$ on machine $m$. 
\end{itemize}
Along with the release times $r_j$, deadlines $d_j$, and weights $w_j$ this constitutes a complete definition of a Job Shop problem. 



To demonstrate the Job Shop scheme, we encode it as an integer linear programming (ILP) problem - that is, the state-of-the-art approach to such a problem. Henceforth, we require some additional definitions:
\begin{itemize}
\item Let $\mathcal{S}_{j,j'}$ be the set of machines used both by $j$ and $j'$. 
\item Let $\sigma(S_j, m)$ be the machine preceding $m$ in the schedule $S_j$. 
\item Let $m_{j, \text{start}}$ and $m_{j, \text{end}}$ be the first and last machine of job $j$. 
\item Let $S^+_j(m)$ be the schedule of job $j$ up to and including machine $m$, and $S^-_j(m)$ the remainder such that concatenating $S^+_j(m)$ and $S^-_j(m)$ produces $S_j$.
\end{itemize}

The ILP encodings employ the decision variables:
\begin{itemize}
    \item $t_{j,m} \in \mathbb{Z}^+$ - time job $j$ is finished on machine $m$,
    \item $y_{j, j', m} \in \{0,1\}$ - precedence variable equal to one if $j$ is performed before $j'$ on $m$.
\end{itemize}

From the definition of a Job Shop scheduling problem, each job must be processed by a series of machines in order.
This requirement yields constraints on the decision variables,
\begin{align}
    t_{j,\sigma(S_j, m)} + p_{j,m} \leq t_{j,m}  
    \qquad \forall {j} \forall {m \in S_j^- (m_{j, \text{start}})}
    .
    \label{eq::ILP_machines_seq}
\end{align}
Additionally, the fact that each machine can only process a single job at a time produces more constraints,
\begin{align}
    t_{j', m} + p_{j,m} &\leq M y_{j,j', m} + t_{j,m} 
    \qquad \forall {j \neq j'}\; \forall {m \in \mathcal{S} {j,j'}}
    .
    \label{eq::ILP_machines_one}
\end{align}
Here, we use the so-called ``big-M encoding'' where $M$ is some number chosen to be large enough that the inequality always holds if $y_{j,j',m} = 1$.
The minimal $M$ can be computed using the minimal value of $t_{j,m}$ and the maximal value of $t_{j',m}$, which we will discuss later.

The release and deadline constraints yield:
\begin{subequations}
\begin{align}
    r_j + p_{j, m_{j, \text{start}}} &\leq t_{j, m_{j, \text{start}}} 
    \qquad \forall j
    ,
    \\
    t_{j, m_{j, \text{end}}} &\leq d_j
    \qquad\qquad \forall j
    .
\end{align}
    \label{eq::ILP_constr_r_d}
\end{subequations}
We can determine the minimum and maximum values of time variables for each (intermediate) machine by:
\begin{subequations}
\begin{align}
    t_{\text{min}}(j,m) &= r_j + \;\smashoperator{\sum_{m' \in S^+_j(m)}}\; p_{j, m'} \leq t_{j, m} 
    \quad \forall j  \forall m \in S_j
    ,
    \\
t_{\text{max}}(j,m) &= d_j - \;\smashoperator{\sum_{m' \in S^-_j(m)}}\; p_{j, m'} \geq t_{j, m} 
    \quad \forall {j} \forall {m \in S_j}
    .
\end{align}
    \label{eq::ILP_constr_limits}
\end{subequations}

Finally, the maximal weighted tardiness objective is represented as,
\begin{align}
    \text{objective} =  \sum_j w'_j t_{j, m_{j, \text{end}}} - \text{offset} 
    ,
    \label{eq::ilp_obj}
\end{align}
where we have introduced 
\begin{subequations}
\begin{align}
    w'_j &=  \frac{w_j}{t_{\max}(j, m_{j, \text{end}}) - t_{\min}(j, m_{j, \text{end}}) }  
    , \label{eq::obj_wprim}
    \\
    \text{offset} &= \sum_j w'_j t_{\min}(j, m_{j, \text{end}}) .
\end{align}
\end{subequations}
Referring to Eq.~\eqref{eq::obj_wprim} the input from each $t_{j, m_{j, \text{end}}}$ does not exceed $w_j$. 

\section{QUBO Encoding}
\label{app:QUBOEncoding}

In this appendix, we describe how the Job Shop problems studied in the main text are transformed into QUBO problems. 
We will use the notation introduced in Appendix~\ref{app:ILPEncoding}, though we do not need to first build an ILP problem to produce a QUBO problem.

One of the key differences which must be overcome in translating a Job Shop scheduling problem into a QUBO is that we must start with a constrained optimization problem and build an equivalent unconstrained optimization problem.
We do this by transforming constraints to penalty terms in the objective.

To enforce the release and deadline constraints of a Job Shop problem on our binary decision variables $\vec{x}$, we must impose the constraint that each job completes exactly once in the interval given in Eq.~\eqref{eq::ILP_constr_limits},
\begin{align}
    ~~\smashoperator{\sum_{t_{\text{min}}(j,m) \leq t \leq t_{\text{max}}(j,m)}}~~ x_{j,m,t} = 1
    \qquad \forall j \forall m \in S_j
    ,
    \label{eq::sum_to_one_constraint}
\end{align}
Notice that this is a constraint on the sum of a collection of the decision variables. 
This may be reformulated in terms of penalty terms and encoded in an unconstrained optimization problem~\cite{lucas2014ising}, which produces terms of the form
\begin{align}
    p_{\rm sum}\sum_{m,j}\pqty{\sum_{t,t': t \neq t'} x_{m,j,t}x_{m,j,t'} - \sum_{t} x_{m,j,t}^2}
    ,
    \label{eq::QUBO_sum}
\end{align}
with the precise choice of indices over which the sums run chosen to match with Eq.~\eqref{eq::sum_to_one_constraint}.

The sequence constraints from Eq.~\eqref{eq::ILP_machines_seq} translate to the decision variables $\vec{x}$ as:
\begin{align}
    \nonumber
    &\forall {j} 
    \forall {m \in S_j^- (m_{j, \text{start}}) } 
    \\\nonumber
    x_{\sigma,j,t'} x_{m, j, t} = 0 
    \qquad
    &\forall {t_{\text{min}}(j,\sigma) \leq t' \leq t_{\text{max}}(j,\sigma)} 
    \\
    &\forall {t_{\text{min}}(j,m) \leq t < t' + p_{j,m} }
    ,
    \label{eq::QUBO_machines_seq}
\end{align}
where to save space we have omitted the arguments to $\sigma$. All occurrences are $\sigma(S_j, m)$.
The constraint that each machine can work on one job at a time from Eq.~\eqref{eq::ILP_machines_one}
translates to:
\begin{align}
    \nonumber
    &\forall {j \neq j'} 
    \forall {m \in \mathcal{S}_{j,j'}} 
    \\\nonumber
    x_{j,m,t} x_{j', m, t'} = 0
    \qquad
    &\forall {-p_{j,m} < t'-t < p_{j',m}}
    \\
    &\forall {t_{\text{min}}(j,m) \leq t \leq t_{\text{max}}(j,m)}  
    ,
    \label{eq::QUBO_machines_one}
\end{align}
where we of course also require $t_{\text{min}}(j',m) \leq t' \leq t_{\text{max}}(j',m)$. 
These constraints along with those from Eq.~\eqref{eq::QUBO_machines_seq} work with pairs of decision variables simultaneously, and just as with the sum constraints they may be transformed into penalty terms suitable for inclusion in a QUBO~\cite{lucas2014ising}.
Doing so produces contributions of the form,
\begin{align}
    p_{\rm pair}\sum_{i<i'}\pqty{x_i x_{i'} + x_{i'} x_i}
    ,
    \label{eq::QUBO_pair}
\end{align}
again with the indices suitable chosen to correspond to Eqs.~\eqref{eq::QUBO_machines_seq} and~\eqref{eq::QUBO_machines_one}.

The objective of Eq.~\eqref{eq::ilp_obj} is implemented as the following:
\begin{align}
    \text{objective}(\vec{x})
    = 
    \sum_j \!\! \sum_{t_{\min} \leq t \leq t_{\max}} \!\! w'_j\,t\,x_{j, m_{j, \text{end}}, t} - \text{offset}
    ,
    \label{eq::QUBO_obj}
\end{align}
where the bounds on $t$ are evaluated as $t_{\min}(j, m_{j, \text{end}})$, $t_{\max}(j, m_{j, \text{end}})$.
To solve the Job Shop problem, this objective function should be minimized subject to the constraints given above.

\bibliography{references}

\end{document}